\documentclass[aps,prd,twocolumn,nofootinbib]{revtex4-1}
\usepackage{epsfig}
\usepackage[colorlinks,linkcolor=blue,anchorcolor=blue,citecolor=blue,urlcolor=blue,breaklinks=true]{hyperref}
\usepackage{graphics}
\usepackage{slashed}
\usepackage{color}
\usepackage{amsmath}
\usepackage{bm}
\usepackage{setspace}
\usepackage{caption}
\usepackage{multirow}

\begin{document}
\author{Ya-Peng Zhao$^{1}$}\email{zhaoyapeng2013@hotmail.com}
\author{Chao-Yong Wang$^{1,2}$}\email{30130808@hncj.edu.cn}
\author{Shu-Yu Zuo$^{3}$}
\author{Cheng-Ming Li$^{4}$}\email{licm@zzu.edu.cn}
\address{$^{1}$ School of Mathematics and Physics, Henan University of Urban Construction, Pingdingshan 467036, China}
\address{$^{2}$ Henan Provincial Engineering Laboratory of Building-Photovoltaics, Pingdingshan 467036, China}
\address{$^{3}$ College of Science, Henan University of Technology, Zhengzhou 450000, China}
\address{$^{4}$ School of Physics and Microelectronics, Zhengzhou University, Zhengzhou 450001, China}
\title{Nonextensive effects on QCD chiral phase diagram and baryon-number fluctuations within Polyakov-Nambu-Jona-Lasinio model}

\begin{abstract}
   In this paper, a version of the Polyakov-Nambu-Jona-Lasinio (PNJL) model based on nonextensive statistical mechanics is presented. This new statistics summarizes all possible factors that violate the assumptions of the Boltzmann-Gibbs (BG) statistics to a dimensionless nonextensivity parameter $q$, and when $q$ tends to 1, it returns to the BG case. Within the nonextensive PNJL model, we found that as $q$ increases, the location of the critical end point (CEP) exhibits non-monotonic behavior. That is, for $q<1.15$, CEP moves in the direction of lower temperature and larger quark chemical potential. But for $q>1.15$, CEP turns to move in the direction of lower temperature and lower quark chemical potential. In addition, we studied the moments of the net-baryon number distribution, that is, the variance ($\sigma^{2}$), skewness (S), and kurtosis ($\kappa$). Our results are generally consistent with the latest experimental data, especially for $\sqrt{S_{NN}}>19.6\ \mathrm{GeV}$, when $q$ is set to $1.07$.
\bigskip

\bigskip

\noindent PACS Number(s): 12.38.Mh, 12.39.-x, 25.75.Nq, 12.38.Aw

\end{abstract}


\maketitle

\section{INTRODUCTION}
Since the discovery of quark-gluon plasma (QGP) at the Relativistic Heavy-Ion Collider (RHIC) and Large Hadron Collider (LHC), the determination of QCD phase diagram at high quark chemical potential, especially the search for the location of critical end point (CEP), has been the focus of experiments, such as the BES-\uppercase\expandafter{\romannumeral2} program at RHIC and the CBM experiment at FAIR.
Theoretically, due to the ``sign problem'' encountered by lattice QCD, people use various effective models to study the QCD phase diagram, such as chiral perturbation theory~\cite{Espriu:2020dge}, finite energy sum rules~\cite{ PhysRevD.84.056004}, Dyson-Schwinger Equations (DSEs)~\cite{Fischer:2018sdj,PhysRevD.91.056003,Zhao_2019}, Nambu-Jona-Lasinio (NJL) model and
Polyakov-Nambu-Jona-Lasinio (PNJL) model~\cite{BUBALLA2005205,Cui:2018bor,PhysRevC.101.065203,PhysRevD.101.096006,ZHAO2020114919}.

It is worth noting that Boltzmann-Gibbs (BG) statistics are often used in the study of QCD phase diagram.
However, in relativistic heavy ion collisions, this is not satisfactory. The collision system is small enough and evolving rapidly that global equilibrium is difficult to establish.
As a result, the probability distribution is no longer the standard $e$-exponential distribution in BG statistics, but becomes the so-called power-law
tailed distribution (i.e. the Tsallis distribution introduced next), which is supported by more and more relativistic heavy ion collision experiments~\cite{Sharma:2019grs,Bhattacharyya:2017cdk,PhysRevC.75.064901,PhysRevC.83.064903,Aamodt2011,Aad:2010ac,Khachatryan2010}.

The most notable difference between Tsallis statistics and BG statistics is that the former replaces the probability distribution as follows~\cite{PhysRevC.77.044903,Eur.Phys.J.A2016,shen2017chiral}:
\begin{eqnarray}\label{Tsallis}
\rho_{BG}(x)=C\mathrm{exp}(x)\longrightarrow \rho_{q}(x)=C_{q}\mathrm{exp}_{q}(x),
\end{eqnarray}
where
\begin{eqnarray}
\mathrm{exp}_{q}(x)=[1+(1-q)x]^{\frac{1}{1-q}},
\end{eqnarray}
and the inverse function is
\begin{eqnarray}
\mathrm{ln}_{q}(x)=\frac{x^{1-q}-1}{1-q}.
\end{eqnarray}
As $q\rightarrow1$, BG statistics is recovered.

In fact, the Tsallis distribution appears not only in relativistic heavy ion collision systems, but also in complex systems with non-ergodicity, multifractality, long-range correlations, long-term
memory, etc~\cite{tirnakli2016the,cirto2018validity,Tsallis:2012js}. Ref.~\cite{tirnakli2016the} shows in a very clear manner that due to a break in ergodicity, the system has crossed from BG statistics to Tsallis statistics.
Furthermore, it should be noted that the research in Ref.~\cite{PhysRevD.101.034019} shows that the renormalizable field theories lead to fractal structures, which can be studied by using Tsallis statistics.
Therefore, in this paper we will use Tsallis statistics to study the QCD phase diagram. And the paper is organized as follows: In Sec. II, we generalize the PNJL model to its nonextensive version. And in Sec. III, we focus on the impact of nonextensive effects on the QCD phase diagram, especially on the location of CEP. In addition, we study the moments of the net-baryon number distribution and compare them with latest experimental data. Finally, we give a brief summary of our work in Sec. IV.
\section{PNJL and nonextensive pnjl model}
\subsection{PNJL model}
The Lagrangian of the two-flavor and three-color PNJL model for equal-mass quark is~\cite{PhysRevD.73.014019,FUKUSHIMA2004277}:
\begin{eqnarray}
\mathcal{L}_{PNJL} &=& \bar{\Psi}(i\gamma_{\mu}D^{\mu}-m)\Psi +G\,[(\bar{\Psi}\Psi)^2+(\bar{\Psi}i\gamma_5\bm{\tau}\Psi)^2]\nonumber\\ &&-\mathcal{U}(\Phi,\bar{\Phi};T),
\label{effective}
\end{eqnarray}
where $m$ is the current quark mass, $\Psi$ denotes the quark field, and G is the four-fermion interaction coupling constant.

The effective Polyakov-loop potential $\mathcal{U}(\Phi,\bar{\Phi};T)$ that accounts for gauge field self-interaction is a function of the Polyakov-Loop $\Phi$ and its Hermitian conjugate $\bar{\Phi}$
\begin{eqnarray}
\Phi=\frac{\langle Tr_{c}L \rangle}{N_{c}}, \ \    \bar{\Phi}=\frac{\langle Tr_{c}L^{\dagger} \rangle}{N_{c}},
\label{phi}
\end{eqnarray}
with the Polyakov line is
\begin{eqnarray}
L(\vec{x})=\mathcal{P}\exp(i\int_{0}^{\beta}A_{4}(\vec{x},\tau)d\tau).
\end{eqnarray}
Following Refs.~\cite{PhysRevD.84.014011,PhysRevD.85.054013,PhysRevD.94.071503,PhysRevD.102.014014}, we take the approximation $L^{\dag}=L$ and have $\Phi=\bar{\Phi}$.

The form of the Polyakov effective potential $\mathcal{U}$ is provided by~\cite{PhysRevD.75.034007},
\begin{eqnarray}
\frac{\mathcal{U_{L}}}{T^{4}}=-\frac{a(T)}{2}\Phi^{2}+b(T)\mathrm{ln}[1-6\Phi^{2}-3\Phi^{4}+8\Phi^{3}],
\end{eqnarray}
with
\begin{eqnarray}
a(T)&=&a_{0}+a_{1}(\frac{T_{0}}{T})+a_{2}(\frac{T_{0}}{T})^{2},\\
b(T)&=&b_{3}(\frac{T_{0}}{T})^{3}.
\end{eqnarray}
The parameters are used to reproduce the pure gluonic lattice data, see Table~\ref{tb2}. Following Ref.~\cite{PhysRevD.82.076003}, we adjust $T_{0}=190\ \mathrm{MeV}$ to take into account the dynamical quarks.

The coupling constant $G$, as pointed out in Refs.~\cite{PhysRevD.82.076003,PhysRevD.82.065024}, should depend on $\Phi$.
Here, we take the form in Ref.~\cite{PhysRevD.82.076003}
\begin{eqnarray}
G=g[1-\alpha_{1}\Phi^{2}-2\alpha_{2}\Phi^{3}].
\end{eqnarray}
The PNJL model with parameters used in Tables~\ref{tb2} and~\ref{tbnjl} is in good agreement with lattice QCD data~\cite{PhysRevD.82.076003}.

The thermodynamic potential density function $\Omega$ is defined as
\begin{eqnarray}
\Omega&=&-\frac{T}{V}\mathrm{ln}Z\\\nonumber
&=&-\frac{T}{V}\mathrm{ln}\mathbf{Tr}\ \mathrm{exp}(-\frac{1}{T}\int d^{3}x(\mathcal{H}-\mu\psi^{\dag}\psi)).
\label{omegaq}
\end{eqnarray}
Using finite temperature field theory and mean field approximation, $\Omega$ can be derived as:
\begin{eqnarray}\label{omega}
\Omega(\mu,T,M,\Phi)&=&\mathcal{U}(\Phi;T)+\frac{(M-m)^{2}}{4G}\\
&-&2N_{c}N_{f}\int_{0}^{\Lambda}\frac{{\rm d}^3\vec{p}}{(2\pi)^3}E_{p}\nonumber\\
&-&2N_{f}T\int_{0}^{\infty}\frac{{\rm d}^3\vec{p}}{(2\pi)^3}(\mathrm{ln}F^{+}+\mathrm{ln}F^{-}),\nonumber
\end{eqnarray}
where $M$ is the dynamical quark mass:
\begin{eqnarray}
M=m-2G(\Phi)\sigma,
\end{eqnarray}
$\sigma=\langle\bar{\Psi}\Psi\rangle$ is the quark chiral condensate. And
\begin{eqnarray}
F^{\pm}&=&1+3\Phi(e^{-(E_{p}\mp\mu)/T}+e^{-2(E_{p}\mp\mu)/T})\nonumber\\
&&+e^{-3(E_{p}\mp\mu)/T},
\end{eqnarray}
in which $E_{p}=\sqrt{p^{2}+M^{2}}$ is the single quasi-particle energy.
The finite temperature contribution term is finite, so here we only impose
the cut-off $\Lambda$ on the vacuum term~\cite{sym2031338,PhysRevD.73.014019,PhysRevC.79.055208,FUKUSHIMA2004277}.
\begin{center}
\begin{table}
\caption{Parameter set used in our work.}\label{tb2}
\begin{tabular}{p{1.4cm} p{1.4cm} p{1.4cm} p{1.4cm} p{1.4cm}}
\hline\hline
$a_0$&$a_1$&$a_2$&$b_3$&$T_{0}(\mathrm{MeV})$\\
\hline
3.51&-2.47&15.2&-1.75&190\\
\hline\hline
\end{tabular}
\end{table}
\end{center}

\begin{center}
\begin{table}
\caption{Parameter set used in our work.}\label{tbnjl}
\begin{tabular}{p{1.5cm} p{2.3cm} p{1.5cm} p{0.9cm} p{0.9cm}}
\hline\hline
$\Lambda(\mathrm{MeV})$&$g(\mathrm{MeV^{-2}})$&$m(\mathrm{MeV})$&$\alpha_{1}$&$\alpha_{2}$\\
\hline
631.5&$5.498\times10^{-6}$&5.5&0.2&0.2\\
\hline\hline
\end{tabular}
\end{table}
\end{center}

Numerical solutions of physical quantities $M$ and $\Phi$ with respect to $\mu$ and $T$ can be obtained by minimizing the thermodynamic potential
\begin{eqnarray}\label{gap}
\frac{\partial\Omega}{\partial M}=\frac{\partial\Omega}{\partial\Phi}=0.
\end{eqnarray}
The quantitative study of the QCD phase transition is based on thermal susceptibility
\begin{eqnarray}\label{kt}
\chi_{T}=\frac{\partial\sigma}{\partial T}.
\end{eqnarray}
\subsection{Nonextensive PNJL model}
The nonextensive version of the thermodynamic potential density function $\Omega_{q}$ is defined as~\cite{2014,shen2017chiral}
\begin{eqnarray}
\Omega_{q}&=&-\frac{T}{V}\mathrm{ln_{q}}Z_{q}\\\nonumber
&=&-\frac{T}{V}\mathrm{ln_{q}}\mathbf{Tr}\ \mathrm{exp}_{q}(-\frac{1}{T}\int d^{3}x(\mathcal{H}-\mu\psi^{\dag}\psi)).
\label{omegaq}
\end{eqnarray}
Combining q-algebra and following the same derivation steps as Eq.~(\ref{omega}), the thermodynamic potential density function $\Omega_{q}$ can be derived as:
\begin{eqnarray}\label{omega00}
\Omega_{q}(\mu,T,M,\Phi)&=&\mathcal{U}(\Phi;T)+\frac{(M-m)^{2}}{4G}\\
&-&2N_{c}N_{f}\int_{0}^{\Lambda}\frac{{\rm d}^3\vec{p}}{(2\pi)^3}E_{p}\nonumber\\
&-&2N_{f}T\int_{0}^{\infty}\frac{{\rm d}^3\vec{p}}{(2\pi)^3}(\mathrm{ln}_{q}F_{q}^{+}+\mathrm{ln}_{q}F_{q}^{-}),\nonumber
\end{eqnarray}
where
\begin{eqnarray}
F_{q}^{\pm}&=&1+3\Phi(e_{q}(-(E_{p}\mp\mu)/T)+e_{q}(-2(E_{p}\mp\mu)/T))\nonumber\\
&&+e_{q}(-3(E_{p}\mp\mu)/T).
\end{eqnarray}
It should be noted that, as a simplification, the form of the Polyakov-loop potential remains unchanged. That is to say, it is only indirectly affected by the nonextensive effects through the saddle point equations. Besides, the parameters of the PNJL model are as usual. Here, we treat $q$ as a thermodynamic variable on the same footing as $T$ and $\mu$. That is to say, we based on the ansatz that the parameters determined at $T=0$, $\mu=0$ and $q=1$ can be used to study the whole region. The same simplifications also appear in Ref.~\cite{shen2017chiral}.

It is worth noting that to ensure that $e_{q}(x)$ is always a non-negative real function, the following condition must be satisfied:
\begin{eqnarray}\label{condition}
[1+(1-q)x]>0.
\end{eqnarray}
During the calculation, we found that the $q$ value cannot be too large, otherwise the CEP position we care about will fall into the non-physical region. Therefore, in this paper we mainly focus on $1 \leq q \leq 1.2$. This is also the typical region of $q$ values found by high-energy collisions.~\cite{CLEYMANS2013351,LI2013352,PhysRevD.91.054025,Azmi_2014}.
In addition, there is no nonextensive effects at low temperature. Because as $T$ tends to zero, $\Omega_{q}$ tends to $\Omega$ as long as $q>1$.

According to Eq.~(\ref{gap}), the numerical results for $M$ and $\Phi$ can be obtained by solving the following nonlinear coupling equations:
\begin{widetext}
\begin{eqnarray}
M&=&m+4GN_{c}N_{f}\int\frac{{\rm d}^3\vec{p}}{(2\pi)^3}\frac{M}{E_{p}}[1-n_{q}-\bar{n}_{q}],\\
0&=&\frac{\partial\mathcal{U}}{\partial\Phi}-\frac{(M-m)^{2}}{4G^{2}}\frac{\partial G}{\partial\Phi}-2N_{c}N_{f}T\int_{0}^{\infty}\frac{{\rm d}^3\vec{p}}{(2\pi)^3}\{\frac{e_{q}(-(E_{p}-\mu)/T)+e_{q}(-2(E_{p}-\mu)/T)}{[1+3\Phi(e_{q}(-(E_{p}-\mu)/T)+e_{q}(-2(E_{p}-\mu)/T))+e_{q}(-3(E_{p}-\mu)/T)]^{q}}\nonumber\\
&&+\frac{e_{q}(-(E_{p}+\mu)/T)+e_{q}(-2(E_{p}+\mu)/T)}{[1+3\Phi(e_{q}(-(E_{p}+\mu)/T)+e_{q}(-2(E_{p}+\mu)/T))+e_{q}(-3(E_{p}+\mu)/T)]^{q}}\},
\end{eqnarray}
where the q-version of the Fermi-Dirac distribution is
\begin{eqnarray}\label{nq}
n_{q}(T,\mu)=\frac{e_{q}^{q}(-3(E_{p}-\mu)/T)+\Phi(e_{q}^{q}(-(E_{p}-\mu)/T)+2e_{q}^{q}(-2(E_{p}-\mu)/T))}{[1+3\Phi(e_{q}(-(E_{p}-\mu)/T)+e_{q}(-2(E_{p}-\mu)/T))+e_{q}(-3(E_{p}-\mu)/T)]^{q}},
\end{eqnarray}
and
\begin{eqnarray}\label{nbarq}
\bar{n}_{q}(T,\mu)=\frac{e_{q}^{q}(-3(E_{p}+\mu)/T)+\Phi(e_{q}^{q}(-(E_{p}+\mu)/T)+2e_{q}^{q}(-2(E_{p}+\mu)/T))}{[1+3\Phi(e_{q}(-(E_{p}+\mu)/T)+e_{q}(-2(E_{p}+\mu)/T))+e_{q}(-3(E_{p}+\mu)/T)]^{q}}.
\end{eqnarray}
\end{widetext}
As expected, for $q\rightarrow1$, the standard distribution function of the usual PNJL model is recovered.

\begin{figure}
\includegraphics[width=0.47\textwidth]{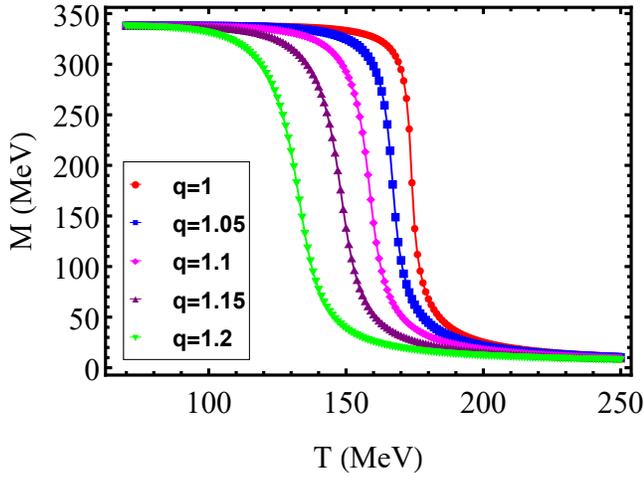}
\caption{Constituent quark mass $M$ as a function of $T$ at $\mu=0$ for five parameters $q$.}
\label{Fig:M}
\end{figure}
\begin{figure}
\includegraphics[width=0.47\textwidth]{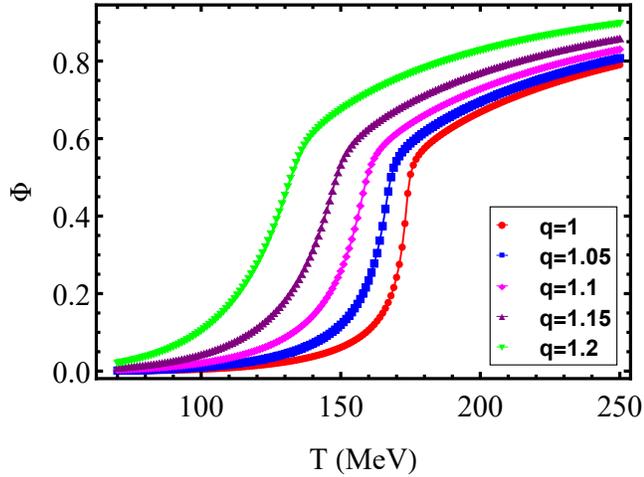}
\caption{Polyakov-loop expectation value $\Phi$ as a function of $T$ at $\mu=0$ for five parameters $q$.}
\label{Fig:phi}
\end{figure}
\section{QCD phase transition and baryon-number susceptibilities within tsallis statistics}
\subsection{QCD phase transition}
The changes of $M$ and $\Phi$ with temperature for different parameters $q$ are shown in Figs.~\ref{Fig:M},~\ref{Fig:phi}. We find that the pseudo-critical temperature decreases with increasing $q$, and the same conclusion can be seen in Ref.~\cite{shen2017chiral}. Besides, even if the nonextensive effects are taken into account, the pseudo-critical temperature of the chiral and deconfinement crossover transition is still the same. For example, at $q=1$ and $q=1.15$, the pseudo-critical temperature is $173\ \mathrm{MeV}$ and $148\ \mathrm{MeV}$, respectively.

\begin{figure}
\includegraphics[width=0.47\textwidth]{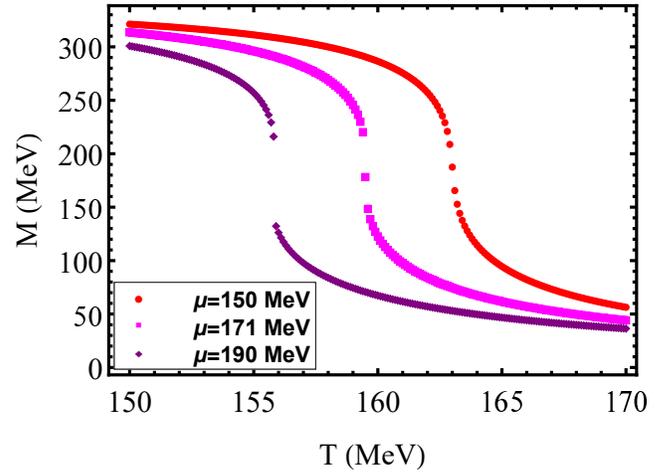}
\caption{Constituent quark mass $M$ as a function of $T$ at $q=1$ for three different quark chemical potentials.}
\label{2}
\end{figure}

\begin{figure}
\includegraphics[width=0.47\textwidth]{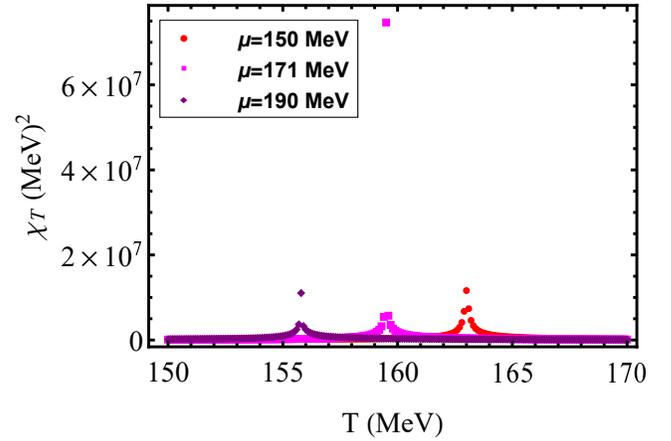}
\caption{The susceptibility $\chi_{T}$ as a function of $T$ at $q=1$ for three different quark chemical potentials.}
\label{1}
\end{figure}

When considering the finite chemical potential, for $q=1$, the variation of the constituent quark mass $M$ with temperature is shown in Fig.~\ref{2}. It can be clearly seen that the first-order phase transition and the crossover transition occur at chemical potentials $\mu=190\ \mathrm{MeV}$ and $\mu=150\ \mathrm{MeV}$, respectively. The determination of the CEP position is based on the peak of the thermal susceptibility $\chi_{T}$. Because in the CEP position, $\chi_{T}$ is divergent. From Fig.~\ref{1}, we determine the location of CEP as $(\mu_{c},T_{c})=(171\ \mathrm{MeV},159.5\ \mathrm{MeV})$.
\begin{figure}
\includegraphics[width=0.47\textwidth]{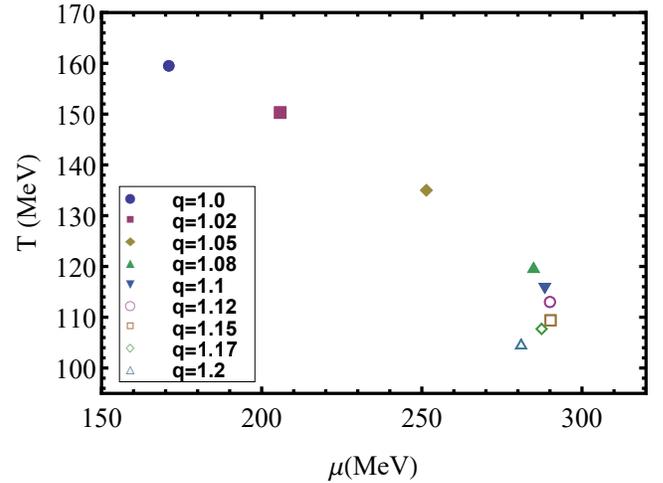}
\caption{The trajectory of CEP position with $q$ in the $T$-$\mu$ plane.}
\label{3}
\end{figure}
\begin{center}
\begin{table}
\caption{Numerical results of CEP for different nonextensivity parameter $q$. The unit of $T_{c}$ and $\mu_{c}$ is MeV.}\label{tb4}
\begin{tabular}{p{0.6cm} p{0.9cm} p{0.9cm} p{0.9cm} p{0.9cm} p{0.7cm} p{0.9cm} p{0.9cm} p{0.8cm} p{0.9cm}}
\hline\hline
$q$&$1$&$1.05$&$1.08$&$1.1$&$1.12$&$1.15$&$1.17$&$1.2$\\
\hline
$T_{c}$&159.5&135&119.8&115.7&113&109.4&107.7&104.7\\
\hline
$\mu_{c}$&171&251.4&284.9&288.4&290&290.2&287.4&281\\
\hline\hline
\end{tabular}
\end{table}
\end{center}

The impact of the nonextensive effects on the location of CEP, which we are most concerned about, is shown in Fig.~\ref{3}. We find that the nonextensive effects have a significant impact on the CEP position. For $1\leq q\leq1.08$, with the increase of $q$, the CEP position moves rapidly in an approximate straight line to the direction of lower temperature and larger chemical potential. But for $1.08\leq q\leq1.2$, we find that the CEP position has an obvious non-monotonic dependence on $q$. Although $T_{c}$ keeps decreasing with the increase of $q$, $\mu_{c}$ first increases and then decreases. The critical value $q_{c}$ is 1.15. The numerical results are shown in Table~\ref{tb4}. This non-monotonic behavior also appeared in our previous article~\cite{Zhao2021}. However, compared with Ref.~\cite{Zhao2021}, the quantitative difference is significant. This is because for $q=1$, the coupling constant $G(\Phi)$ doubles the value of $T_{c}$, and the nonextensive effects become more pronounced only for sufficiently high temperatures.
In addition, another difference is that in Ref.~\cite{Zhao2021}, as $q$ increase, $T_{c}$ also moves in the opposite direction, and the corresponding critical value $q$ is 1.1.
The reappearance of this non-monotonic behavior deserves our attention, and it may be model-independent. The reason why Refs.~\cite{Eur.Phys.J.A2016,shen2017chiral} did not find this phenomenon is most likely because they only calculated $q$ to 1.1.
Our results are meaningful for search of the CEP position in the relativistic heavy-ion collision, where strong fluctuations and long-range correlation drag the system into out-of-equilibrium states, and BG statistics fails. For a wide class of such systems, it has been shown in recent years that the correct approach is to use Tsallis statistics instead~\cite{tirnakli2016the,cirto2018validity}. We find that in the search for CEP, the nonextensive effects cannot be ignored. And the search for CEP in larger or smaller quark chemical potential regions depends on $q_{c}$.
\subsection{Baryon-number susceptibilities}
In this chapter, we pay attention to the baryon-number susceptibilities because they are related to the moments of the conserved net-baryon number distribution, such as the variance $\sigma^{2}$, the skewness $S$, and the kurtosis $\kappa$. Non-monotonic dependence of these moments, especially high-order quantities, on collision energy $\sqrt{S_{NN}}$ is suggested as an experimental signature of the QCD critical point~\cite{Stephanov:2008qz}.
The detailed formulas are as follows:
\begin{eqnarray}
S\sigma&=&\frac{T\chi_{B}^{(3)}}{\chi_{B}^{(2)}},\nonumber\\
\kappa\sigma^{2}&=&\frac{T^{2}\chi_{B}^{(4)}}{\chi_{B}^{(2)}}.
\end{eqnarray}
Where the nth order of the baryon-number susceptibility, $\chi_{B}^{(n)}$ is
\begin{eqnarray}
\chi_{B}^{(n)}=\frac{\partial^{n-1}}{\partial\mu_{B}^{n-1}}\rho_{B}=\frac{\partial^{n-1}}{3^{n}\partial\mu^{n-1}}\rho(T,\mu).
\end{eqnarray}
According to Eq.~(\ref{omega00}), the quark number density $\rho(T,\mu)$ is
\begin{eqnarray}
\rho(T,\mu)&=&-\frac{\partial\Omega_{q}}{\partial\mu}\nonumber\\
&=&2N_{c}N_{f}\int\frac{d^{3}p}{(2\pi)^{3}}(n_{q}(T,\mu)-\bar{n}_{q}(T,\mu)).
\end{eqnarray}
Here for brevity, we omit the subscript $q$ of the quark number density.
\begin{figure}
\includegraphics[width=0.47\textwidth]{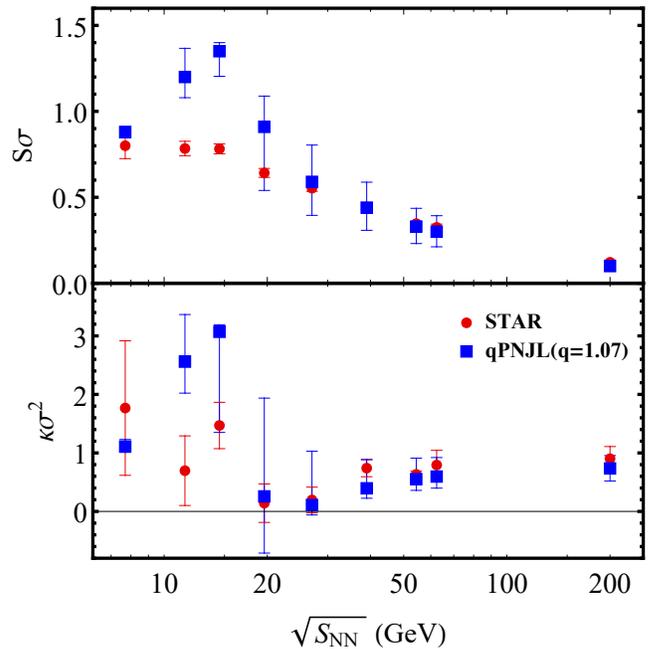}
\caption{Comparison of the qPNJL model results and
the latest experimental data~\cite{STAR:2020tga} for $S\sigma$, $\kappa\sigma^{2}$ at $\sqrt{S_{NN}}=7.7, 11.5, 14.5, 19.6, 27, 39, 54.4, 62.4$, and $200\ \mathrm{GeV}$. The red circles are the experimental data and our qPNJL model results are shown by the blue squares.}
\label{4}
\end{figure}
\begin{figure}
\includegraphics[width=0.47\textwidth]{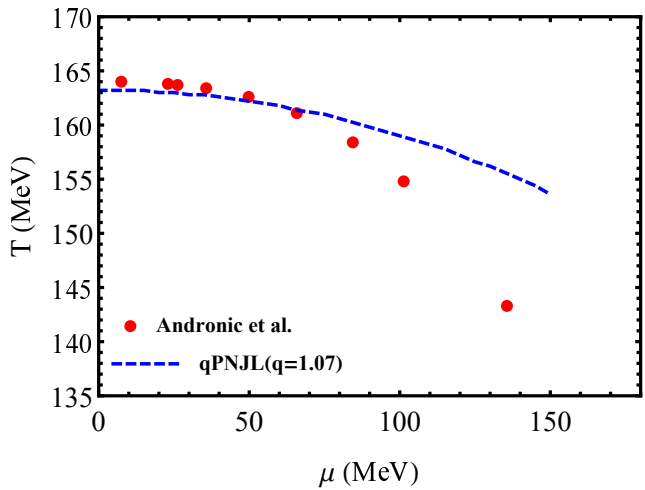}
\caption{The dotted blue line represents the crossover transition line. The red circles represent the chemical freeze-out temperature as a function of the quark chemical potential at $\sqrt{S_{NN}}=7.7, 11.5, 14.5, 19.6, 27, 39, 54.4, 62.4$, and $200\ \mathrm{GeV}$. The intersection of the two is just around $\sqrt{S_{NN}}= 20\ \mathrm{GeV}$.}
\label{5}
\end{figure}

In order to compare with the experiment, one
needs the values of $T$ and $\mu_{B}$ along the freeze-out curve as a function of the collision energy $\sqrt{S_{NN}}$. Here we use the empirical fitting formula in Ref.~\cite{Andronic:2009jd}, as follows
\begin{eqnarray}
T&=&T_{lim}\frac{1}{1+exp(2.60-ln(\sqrt{S_{NN}(GeV)})/0.45)},\nonumber\\
\mu_{B}&=&\frac{1303}{1+0.286\sqrt{S_{NN}(GeV)}},
\end{eqnarray}
with the limiting temperature $T_{lim}=164\ \mathrm{MeV}$.

In Fig.~\ref{4}, $S\sigma$, $\kappa\sigma^{2}$ are shown as a function of $\sqrt{S_{NN}}$.
The latest experimental data come from $Au+Au$ collisions at RHIC~\cite{STAR:2020tga}. By changing the $q$ value ($1\leq q\leq1.2$, $\Delta q=0.01$), it is not difficult to find that our results are generally consistent with the experimental data at $q=1.07$, especially for $\sqrt{S_{NN}}>19.6\ \mathrm{GeV}$, which fits the experimental data well. But for $\sqrt{S_{NN}}=11.5, 14.5\ \mathrm{GeV}$, there is a larger deviation.
The results of $\kappa\sigma^{2}$ are qualitatively consistent with Ref.~\cite{Bzdak:2019pkr}. This non-monotonic variation with $\sqrt{S_{NN}}$ shows that the chemical freeze-out points cross the crossover transition line, see Fig.~\ref{5}. And the intersection of the two is about $\sqrt{S_{NN}}=20\ \mathrm{GeV}$. However, it should be noted that based on our model results, $S\sigma$ also has a non-monotonic dependence on the collision energy, which is qualitatively different from the experimental data and deserves further study. This non-monotonic behavior is also seen in the NJL model~\cite{Fan:2019gkf}.
In addition, we also test the dependence of our results on the $T$, $\mu$, taking $T(\sqrt{S_{NN}})\pm1\ \mathrm{MeV}$ and $\mu(\sqrt{S_{NN}})\pm4\ \mathrm{MeV}$. From Figs.~\ref{4},~\ref{5}, it can be found that near the intersection, there is a large uncertainty in the results. This is due to a large uncertainty of quark-number susceptibility in this region.
All in all, our results generally agree with the experimental data at $q=1.07$, which to some extent indicates that the nonextensive effects are worth considering in relativistic heavy ion collisions. And it may also provide some useful clues for related research to interpret experimental data from a nonextensive perspective.
\section{Summary and Conclusion}
In this paper, combined with the Tsallis statistics and the PNJL model, first of all, we investigated the sensitivity of the QCD phase transition to deviations from usual BG statistics. At zero chemical potential and finite temperature, we found that the pseudo-critical temperature of the chiral and deconfinement transition decreases as $q$ increases. At finite chemical potential and finite temperature, the most interesting thing is that we found the non-monotonic dependence of CEP on $q$. At the beginning, as $q$ increases, CEP moves rapidly toward a lower temperature and larger quark chemical potential. But when $q\geq 1.15$, CEP turns to move in a direction with lower temperature and quark chemical potential. This means that searching for CEP in larger or smaller quark chemical potential regions in relativistic heavy-ion collisions depends on a critical value $q_{c}$.

Secondly, we studied the moments of the net-baryon number
distribution and obtained that $S\sigma$, $\kappa\sigma^{2}$ are generally consistent with the experimental data when $q=1.07$. This indicates that the nonextensive effects may be worth considering in relativistic heavy-ion collisions. In addition, quark stars, as candidates for observed massive stars ($\geq2M_{\odot}$), have attracted much attention in astronomy~\cite{Li:2019ztm,Chu:2019ipr,Li:2019akk,Chen:2016ran}. Therefore, studying the influence of nonextensive effects on the structure and evolution of protoquark stars is a very meaningful topic~\cite{Lavagno:2011zm}.

Finally, how to introduce the nonextensive effects directly into the pure gluonic sector is a problem worthy of further study. As mentioned above, in the standard Polyakov-loop potential, there is no space to introduce nonextensive effects because this potential is based on group integral rather than momentum integral. Can we try to consider other gluonic potentials, for example the Meisinger-Miller-Ogilvie model~\cite{Meisinger:2001cq}, which can describe the confinement-deconfinement nature of QCD, qualitatively. Of course, this will be a brand new attempt, and these issues are our future research directions.
\acknowledgments
This work is supported by the National Natural Science Foundation of China (under Grants No. 12005192), and the Project funded by China Postdoctoral Science Foundation (Grant No. 2020M672255 and No. 2020TQ0287).
\bibliography{cpc}
\end{document}